\documentclass[runningheads,a4paper]{llncs}
\pdfoutput=1
\usepackage{amssymb}
\usepackage{amsmath}
\usepackage{graphicx}
\usepackage{url}
\usepackage{adjustbox}
\usepackage{array}
\usepackage[labelformat=empty]{subfig}
\usepackage[hyperindex,pdftex]{hyperref}
\usepackage[colorinlistoftodos]{todonotes}
\usepackage{booktabs} 
\usepackage{rotating}
\usepackage{multirow}
\usepackage{transparent}
\usepackage{tablefootnote}

\usepackage[subtle,margins=normal,leading=normal]{savetrees} 
\usepackage{bm} 
\usepackage{colortbl}
\usepackage{tabulary}

\begin{document}
\mainmatter              

\title{Cardiac Segmentation of LGE MRI\\ with Noisy Labels}
\titlerunning{Cardiac Segmentation of LGE MRI with Noisy Labels}

\author{
Holger Roth \and Wentao Zhu \and Dong Yang \and Ziyue Xu \and Daguang Xu
}

\institute{
NVIDIA\thanks{Contact: \texttt{\{hroth,wentaoz,dongy,ziyuex,daguangx\}@nvidia.com}}
}
\authorrunning{Holger Roth et al.}

\maketitle              

\begin{abstract}
In this work, we attempt the segmentation of cardiac structures in late gadolinium-enhanced (LGE) magnetic resonance images (MRI) using only minimal supervision in a two-step approach. 
In the first step, we register a small set of five LGE cardiac magnetic resonance (CMR) images with ground truth labels to a set of 40 target LGE CMR images without annotation. Each manually annotated ground truth provides labels of the myocardium and the left ventricle (LV) and right ventricle (RV) cavities, which are used as atlases. After multi-atlas label fusion by majority voting, we possess noisy labels for each of the targeted LGE images. A second set of manual labels exists for 30 patients of the target LGE CMR images, but are annotated on different MRI sequences (bSSFP and T2-weighted). Again, we use multi-atlas label fusion with a consistency constraint to further refine our noisy labels if additional annotations in other modalities are available for a given patient.
In the second step, we train a deep convolutional network for semantic segmentation on the target data while using data augmentation techniques to avoid over-fitting to the noisy labels. After inference and simple post-processing, we achieve our final segmentation for the targeted LGE CMR images, resulting in an average Dice of 0.890, 0.780, and 0.844 for LV cavity, LV myocardium, and RV cavity, respectively.
\keywords{\footnotesize LGE MRI, CMR, Cardiac Segmentation, Deep Learning, Multi-Atlas Label Fusion, Noisy Labels}
\end{abstract}
\section{Introduction}
\label{sec:intro}
\begin{figure*}[htbp!]
	\centering
	\footnotesize
	\begin{tabular}{ccc}
		\hfill
		\hfill
		\subfloat[(a) bSSFP]{\adjincludegraphics[valign=c,height=1.4cm]{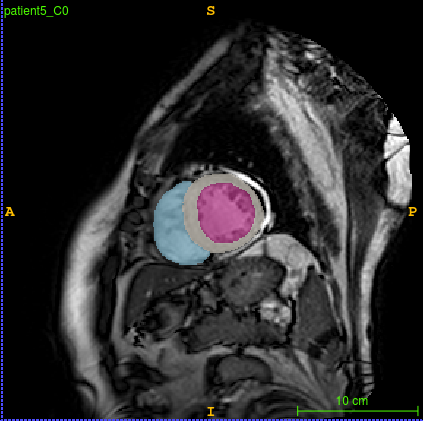}} &  
		\hfill
		\subfloat[(b) T2]{\adjincludegraphics[valign=c,height=1.4cm]{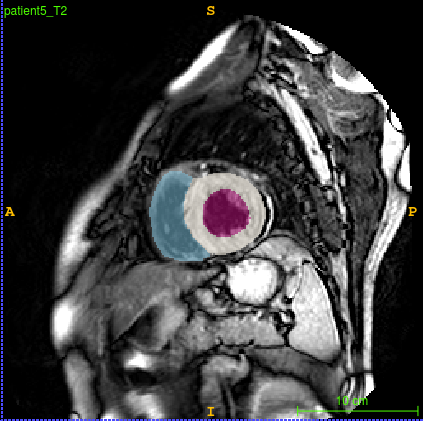}} & 
		\hfill
		\subfloat[(c) LGE]{\adjincludegraphics[valign=c,height=1.4cm]{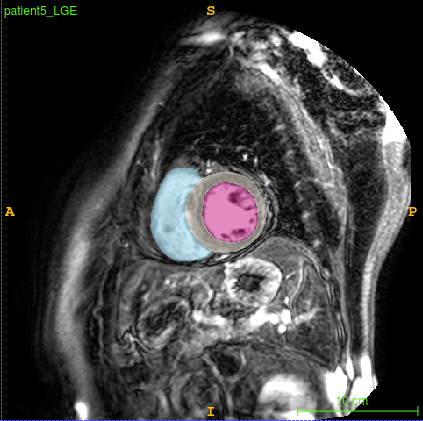}}
	\end{tabular}
	\caption{\footnotesize Sagittal view of different cardiac magnetic resonance (CMR) image sequences of the same patient's heart. Images (a-c) show balanced-steady state free precession (bSSFP), T2-weighted, and late gadolinium-enhanced (LGE) images with overlays of the corresponding manual ground truth (g.t.) annotations [patient 2 of the challenge dataset].
	\label{fig:data}}
\end{figure*}

Segmentation of cardiac structures in magnetic resonance images (MRI) has potential uses for many clinical applications. In particular for cardiac magnetic resonance (CMR) images, late gadolinium-enhanced (LGE) imaging is useful to visualize and detect myocardial infarction (MI). Another common CMR sequence is T2-weighted imaging which highlights acute injury and ischemic regions. Additionally, balanced-steady state free precession (bSSFP) cine sequences can be utilized to analyze the cardiac motion of the heart \cite{zhuang2018multivariate,zhuang2016multivariate}. 
Each CMR sequence is typically acquired independently, and they can exhibit significant spatial deformations among each other even when stemming from the same patient. Nevertheless, segmentation of different anatomies from LGE could still benefit from the combination with the other two sequences (T2 and bSSFP) and their annotations. An example of different CMR sequences utilized in this work can be seen in Fig. \ref{fig:data}. 
LGE enhances infarcted tissues in the myocardium and therefore is an important sequence to focus on for the detection and quantification of myocardial infarction. The infarcted myocardium tissue appears with a distinctively brighter intensity than the surrounding healthy regions. In particular, LGE images are important to estimate the extent of the infarct in comparison to the myocardium \cite{zhuang2018multivariate}. However, manual delineation of the myocardium is time-consuming and error-prone. Therefore, automated and robust methods for providing a segmentation of the cardiac anatomy around the left ventricle (LV) are needed to support the analysis of myocardial infarction.
Modern semantic segmentation methods utilizing deep learning have significantly improved the performance in various medical imaging applications \cite{cciccek20163d,milletari2016v,myronenko20183d,zhu2019anatomynet}. At the same time, deep learning methods typically require large amounts of annotated data in order to train sufficiently robust and accurate models depending on the difficulty of the task. However, in many use cases, the availability of such annotated cases may be limited for a specific targeted image modality or sequence. For CMR applications containing multiple sequences, annotations for the same anatomy of interest might be available for sequences other than the target one of the same patient.
In this work, we attempt the segmentation of cardiac structures in LGE cardiac magnetic resonance (CMR) images utilizing classical methods from multi-atlas label fusion in order to provide ``noisy'' pseudo labels to be used for training deep convolutional neural network segmentation models. 
%

%
\def \Tforward {\mathbf{T}_\mathrm{forward}}
\def \Tbackward {\mathbf{T}_\mathrm{backward}}
\section{Method}
\label{sec:methods}
Our method can be described in two steps. In the first step, we register a small set, e.g. 5, LGE CMR with ground truth labels (``atlases'') to a set of target LGE CMR images without annotation. Each ground truth atlas provides manually annotated labels of the myocardium, and the left and right ventricle cavities. After multi-atlas label fusion by majority voting, we possess noisy labels for each of the targeted LGE images. A second set of manual labels exists for some of the patients of the targeted LGE CMR images, but are annotated on different MRI sequences (bSSFP and T2-weighted). Again, we use multi-atlas label fusion with a consistency constraint to further refine our noisy labels if additional annotations in other sequences are available for that patient.
In the second step, we train a deep convolutional network for semantic segmentation on the target data while using data augmentation techniques to avoid over-fitting to the noisy labels. After inference and simple post-processing, we arrive at our final label for the targeted LGE CMR images.
\subsection{Multi-Atlas Label Fusion of CMR}
\label{sec:label_fusion}
Many methods of multi-atlas label fusion exist \cite{iglesias2015multi}. In this work, we use a well-established non-rigid registration framework based on a B-spline deformation model \cite{Rueckert1999nonrigid} using the implementation provided by \cite{modat2010fast}. The registration is driven by a similarity measurement $\mathcal{S}$ based on intensities from LGE, T2, and bSSFP images.
We perform two sets of registrations
\begin{enumerate}
    \footnotesize
    \item Inter-patient and intra-modality registration, i.e. the registration of LGE with annotations to the targeted LGE images of different patients.
    \item Intra-patient and inter-modality registration, i.e. the registration of bSSFP/T2 with annotations to the targeted LGE images of the same patient.
\end{enumerate}
In both cases, an initial affine registration is performed followed by non-rigid registration between the source image $F$ (providing annotation, i.e. the ``atlas'') and the targeted reference image $R$. A coarse-to-fine registration scheme is used in order to first capture large deformations between the images, followed by more detailed refinements. The deformation is modeled with a 3D cubic B-spline model using a lattice of control points $\{\vec{\phi}\}$ and spacings between the control points of $\delta_x$, $\delta_y$, and $\delta_z$ along the $x$-, $y$-, and $z$-axis of the image, respectively.
Hence, the deformation $\mathbf{T}(\vec{x})$ of a voxel $\vec{x}=(x, y, z)$ to the domain $\Omega$ of the target image can be formulated as
\begin{equation}
\footnotesize
  \label{equ:T_3D}
	\mathbf{T}(\vec{x}) = \sum_{i,j,k}{ \beta^3(\frac{x}{\delta_x} - i) \times \beta^3(\frac{y}{\delta_y} - j) \times \beta^3(\frac{y}{\delta_z} - k) \times \vec{\phi}_{ijk} }.
\end{equation} 
Here, $\beta^3$ represents the cubic B-Spline function.
By maximizing an overall objective function
\begin{multline} 
\footnotesize
	\mathcal{O}\left(I_\mathrm{p},I_\mathrm{s}\left(\mathbf{T}\right);\{\vec{\phi}\}\right) = 	\left(1 - \alpha - \beta\right)\times\mathcal{S} 
	- {\alpha}\times\mathcal{C}_\mathrm{smooth}(\mathbf{T}) -{\beta}\times\mathcal{C}_\mathrm{inconsistency}(\mathbf{T}),
\end{multline}
we can find the optimal deformation field between source and targeted images.
Here, the similarity measure $\mathcal{S}$ is constrained by two penalties $\mathcal{C}_\mathrm{smooth}$ and $\mathcal{C}_\mathrm{inconsistency}$ which aim to enforce physically plausible deformations. The contribution of each penalty term can be controlled with the weights $\alpha$ and $\beta$, respectively.
We use normalized mutual information (NMI) \cite{studholme1999overlap} which is commonly used in inter-modality registrations \cite{iglesias2015atlas} as our driving similarity measure
\begin{equation} 
\footnotesize
	\mathcal{S} = \frac{   H(R)  + H(F(\mathbf{T}))  }{   H(R, F(\mathbf{T})) }.
\end{equation} 
Here, $H(R)$ and $H(F(\mathbf{T}))$ are the two marginal entropies, and $H(R, F(\mathbf{T})$) is the joint entropy. In \cite{modat2010fast}, a Parzen Window (PW) approach \cite{mattes2003pet} is utilized to fill the joint histogram necessary in order to compute the NMI between the images efficiently.
To encourage realistic deformations, we utilize bending energy which controls the ``smoothness'' of the deformation field across the image domain $\Omega$:
\begin{multline} 
\footnotesize
	\mathcal{C}_\mathrm{smooth} =
	 \frac{1}{N} \sum_{\vec{x}\in\Omega}(
	 \left|\frac{\partial^2\mathbf{T}\left(\vec{x}\right)}{{\partial}x^2}\right|^2 +
        \left|\frac{\partial^2\mathbf{T}\left(\vec{x}\right)}{{\partial}y^2}\right|^2 + 
        \left|\frac{\partial^2\mathbf{T}\left(\vec{x}\right)}{{\partial}z^2}\right|^2 \\
+ 2\times\left[\left|\frac{\partial^2\mathbf{T}\left(\vec{x}\right)}{{\partial}xy}\right|^2 +
                \left|\frac{\partial^2\mathbf{T}\left(\vec{x}\right)}{{\partial}yz}\right|^2 +
                \left|\frac{\partial^2\mathbf{T}\left(\vec{x}\right)}{{\partial}xz}\right|^2 
     \right])
.
\end{multline} 
In an ideal registration, the optimized transformations from $F$ to $R$ (forward) and $R$ to $F$ (backward) are the inverse of each other. i.e. $\Tforward=\Tbackward^{-1}$ and $\Tbackward=\Tforward^{-1}$ \cite{feng2009new}. The used implementation by \cite{modat2012inverse} follows the approach by \cite{feng2009new} using compositions of $\Tforward$ and $\Tbackward$ in order to include a penalty term that encourages inverse consistency of both transformations: 
\begin{multline}
\footnotesize
	\label{inv_const_penalty}
	\mathcal{C}_\mathrm{inconsistency} = 
		\sum_{\vec{x}\in \Omega}\left\|\Tforward\left(\Tbackward\left(\vec{x}\right)\right)\right\|^2
	+ \sum_{\vec{x}\in \Omega}\left\|\Tbackward\left(\Tforward\left(\vec{x}\right)\right)\right\|^2
\end{multline}
At each level of the registration, both the image and control point grid resolutions are doubled compared to the previous level. We find suitable registration parameters for both type 1) and type 2) registrations using visual inspection of the transformed image and ground truth atlases.
For type 1) registrations, multiple atlases are available to be registered with each target image. We perform a simple majority voting in order to generate our ``noisy'' segmentation label $\hat{Y}$ for each target image $X$.
\subsection{Label Consistency with Same Patient Atlases}
Because of anatomical consistency between different sequences of the same patient, we employ inter-modality registration to obtain noisy labels for LGE images in type 2) registrations. Two sets of segmentations, denoted by $\hat{Y}_{bSSFP}^{LGE}$ and $\hat{Y}_{T2}^{LGE}$, can be obtained from the registrations: bSSFP to LGE, and T2 to LGE. In order to make sure our noisy labels are accurate enough, we only employ the consistent region $\hat{Y}_{bSSFP}^{LGE} \bigcap \hat{Y}_{T2}^{LGE}$ where both segmentations agree. In the non-consistent regions, we still use the noisy label from type 1) registrations.
In type 1) registrations, we use symmetric registration with bending energy factor $\alpha=0.001$ and inconsistency factor $\beta=0.001$. We use five resolution levels and the maximal number of iteration per level is $300$. The final grid spacing along $x$, $y$ and $z$ are the same with five voxels. In type 2) registrations, we use six levels and the maximal number of iteration per level is $4000$. The final grid spacing along $x$, $y$ and $z$ are the same with one voxel.
       
%
%
\subsection{Deep Learning based Segmentation with Noisy Labels}
In the second step, we train different deep convolutional networks for semantic segmentation on the target data while using data augmentation techniques (rotation, scaling, adding noise, etc.) to avoid over-fitting to the noisy labels. 

Given all pairs of images $X$ and pseudo labels $\hat{Y}$, we re-sample them to 1 $mm^3$ isotropic resolution and train an ensemble $\mathcal{E}$ of $n$ fully convolutional neural networks to segment the given foreground classes, with $P(X) = \mathcal{E}(X)$ standing for the \textit{softmax} output probability maps for the different classes in the image. 
Our network architectures follow the encoder-decoder network proposed in \cite{liu20183d}, named \textit{AH-Net}, and \cite{myronenko20183d} based on the popular 3D U-Net architecture \cite{cciccek20163d} with residual connections \cite{he2016deep}, named \textit{SegResNet}. For training and implementing these neural networks, we used the \textit{NVIDIA Clara Train SDK}\footnote{\url{https://devblogs.nvidia.com/annotate-adapt-model-medical-imaging-clara-train-sdk}} and NVIDIA Tesla V100 GPU with 16 GB memory.
As in \cite{liu20183d}, we initialize \textit{AH-Net} from \textit{ImageNet} pretrained weights using a ResNet-18 encoder branch, utilizing anisotropic ($3 \times 3 \times 1$) kernels in the encoder path in order to make use of pretrained weights from 2D computer vision tasks. 
While the initial weights are learned from 2D, all convolutions are still applied in a full 3D fashion throughout the network, allowing it to efficiently learn 3D features from the image. In order to encourage view differences in our ensemble models, we initialize the weights in all three major 3D image planes, i.e. $3 \times 3 \times 1$, $3 \times 1 \times 3$, and $1 \times 3 \times 3$, corresponding to axial, sagittal, and coronal planes of the images. This approach results in three distinct \textit{AH-Net} models to be used in our ensemble $\mathcal{E}$.
The Dice loss \cite{milletari2016v} has been established as the objective function of choice for medical image segmentation tasks. Its properties make it suitable for the unbalanced class labels common in 3D medical images:
\begin{equation}
    \mathcal{L}_{Dice} = 1 - \frac{2\sum_{i=1}^{N}y_i\hat{y}_i}{\sum_{i=1}^{N} y_i^2 + \sum_{i=1}^{N} \hat{y}_i^2}
    \label{eq:dice}
\end{equation}
Here, $y_i$ is the predicted probability from our network $f$ and $\hat{y_i}$ is the label from our ``noisy'' label map $\hat{Y}$ at voxel $i$. For simplicity we show the Dice loss for one foreground class in Eq. \ref{eq:dice}. In practice, we minimize the average Dice loss across the different foreground classes.
After inference and simple post-processing, we arrive at our final label set for the targeted LGE CMR images. We resize the ensemble models' prediction maps to the original image resolution using trilinear interpolation, fuse each probability map using an \textit{median} operator in order to reduce outliers. Then, the label index is assigned using the \textit{argmax} operator:
\begin{equation}
    Y(X) = \operatorname*{argmax}\left(\ \operatorname*{median}\left(\ \left\{\mathcal{E}_0(X), \dots, \mathcal{E}_n(X)\right\} \ \right) \ \right)
    \label{eq:fusion}
\end{equation}
Finally, we apply 3D largest connected component analysis on the foreground in order to remove isolated outliers.

\section{Experiments \& Results}
\label{sec:experiments}
\subsection{Challenge Data}
\begin{figure*}[htbp!]
	\centering
	\footnotesize
	\begin{tabular}{ccccc}
		\subfloat[(a) LGE]{\adjincludegraphics[valign=c,height=1.1cm]{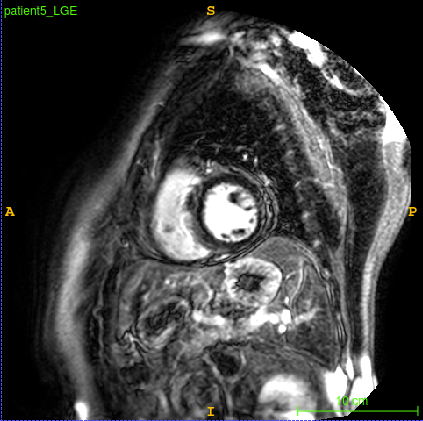}} &
		\hfill
		\subfloat[(b) g.t.]{\adjincludegraphics[valign=c,height=1.1cm]{figs/patient5_LGE_gt}} &
		\hfill
		\subfloat[(c) g.t. 3D]{\adjincludegraphics[valign=c,height=1.1cm]{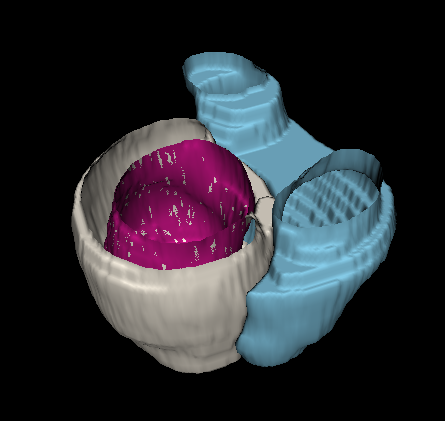}} & 
		\hfill
		\subfloat[(d) pred.]{\adjincludegraphics[valign=c,height=1.1cm]{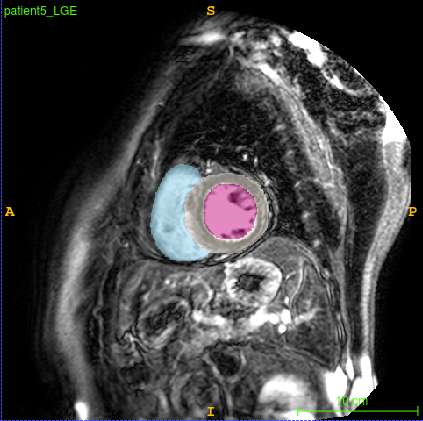}} &
		\hfill
		\subfloat[(e) pred. 3D]{\adjincludegraphics[valign=c,height=1.1cm]{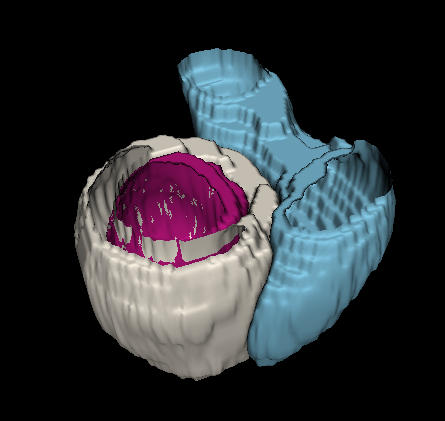}}
	\end{tabular}
	\caption{\footnotesize Comparison of the available ground truth annotation (b) and (c) in a validation LGE dataset and our model's prediction (d) and (e) [patient 2 of the challenge dataset]. \label{fig:validation_result}}
\end{figure*}

The challenge organizers provided the anonymized imaging data of 45 patients with cardiomyopathy who underwent CMR imaging at the Shanghai Renji hospital, China, with institutional ethics approval. For each patient, three CMR sequences (LGE, T2, and bSSF) are provided as multi-slice images in the ventricular short-axis views acquired at breath-hold.  
Slice-by-slice manual annotations of the right and left ventricular, and ventricular myocardium have been generated as gold-standard using ITK-SNAP\footnote{\url{http://www.itksnap.org}} for training of the mdoels and for evaluation the segmentation results. The manual segmentation took about 20 minutes/case as stated by the challenge organizers. We also use ITK-SNAP for all the visualizations shown in this paper.
For more details, see the challenge website\footnote{\url{https://zmiclab.github.io/mscmrseg19/data.html}}. The available training and test data have the following characteristics:
\begin{multicols}{2}
\paragraph{Training data:} 
\footnotesize
\begin{itemize}
    \item Patient 1-5: 
    \begin{itemize}
        \item LGE CMR (image + manual label) for validation
        \item T2-weighted CMR (image + manual label)
        \item bSSFP CMR (image + manual label)
    \end{itemize}    
    \item Patient 6-35:
    \begin{itemize}
        \item T2-weighted CMR (image + manual label)
        \item bSSFP CMR (image + manual label) 
    \end{itemize}
    \item Patient 36-45:
    \begin{itemize}
        \item T2-weighted CMR (only image)
        \item bSSFP CMR (only image)
    \end{itemize}
\end{itemize}
\vfill\null
\columnbreak
\paragraph{Test data:} 
\footnotesize
\begin{itemize} 
    \item Patient 6-45:
    \begin{itemize}
        \item LGE CMR (only image)
    \end{itemize}
\end{itemize}
\end{multicols}
\noindent As one can see, only five ground truth annotations are available in the targeted LGE images. However, 30 images have gold standard annotations available in different image modalities, i.e. bSSFP and T2. We use all available annotations for type 1) and type 2) multi-atlas label fusion approaches described in Section \ref{sec:methods}. After ``noisy'' label generation for all testing LGE images, we train our deep neural network ensemble to produce the final prediction labels for 40 LGE images in the test set. The five manually annotated LGE cases are used as the validation set during deep neural network training in order to find the best model parameters and avoid overfitting completely to the noisy labels. Throughout the challenge, the authors are blinded to the ground truth of the test set during model development and evaluation.
Our evaluation scores on the test set are summarized in Table \ref{tab:results}. A comparison of the available ground truth annotation in a validation LGE dataset and our model's prediction is shown in Fig. \ref{fig:validation_result}.
%
\begin{table}[htbp!]
    \caption{\footnotesize Evaluation scores on 40 LGE test images as provided by the challenge organizers. Both overlap and surface distance-based metrics are shown. LV and RV denote the left and right ventricle, respectively.}
      \centering
      \footnotesize
\begin{tabular}{|c|c|c|c|c|}    
\hline
\textbf{Metric} & \textbf{LV Cavity} & \textbf{LV Myocardium}	& \textbf{RV Cavity} & \textbf{Average}\\
\hline\hline
Dice                    & 0.890 &	0.780 &	0.844 &	0.838 \\
\hline
Jaccard                 & 0.805 &	0.642 &	0.735 &	0.727 \\
\hline
Surface distance [mm]   & 2.13 &	2.32 &	2.80 &	2.41 \\
\hline
Hausdorff distance [mm] & 11.6 &	16.3 &	18.1 &	15.3 \\
\hline
\end{tabular}\label{tab:results}
\end{table}
%
%
\section{Discussion \& Conclusion}
\label{sec:conclusions}
\begin{figure*}[htbp!]
	\centering
	\footnotesize
	\begin{tabular}{ccccc}
		\subfloat[(a) LGE]{\adjincludegraphics[valign=c,height=1.1cm]{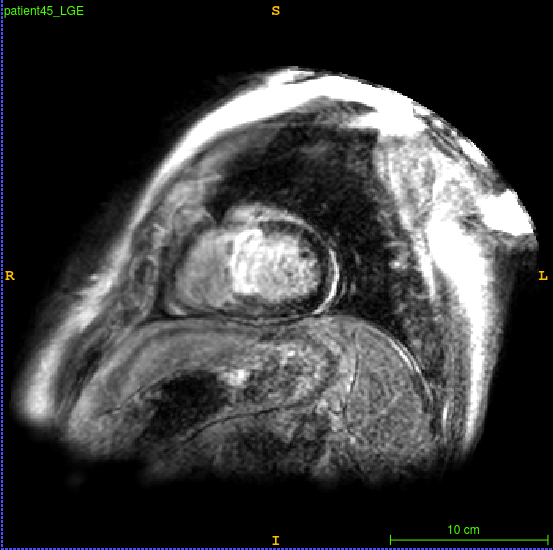}} &
		\hfill
		\subfloat[(b) pseudo]{\adjincludegraphics[valign=c,height=1.1cm]{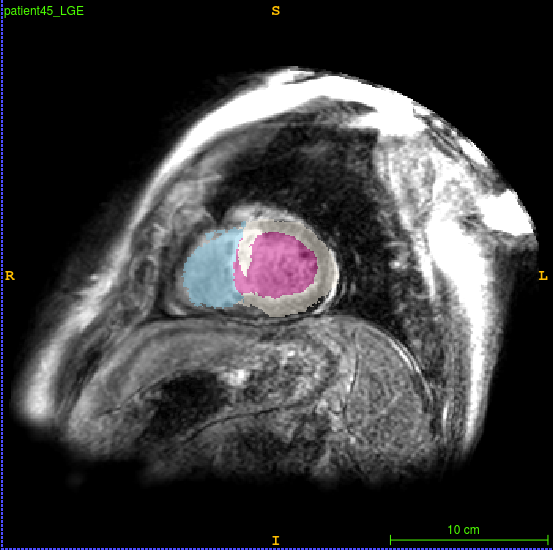}} &
		\hfill
		\subfloat[(c) pseudo]{\adjincludegraphics[valign=c,height=1.1cm]{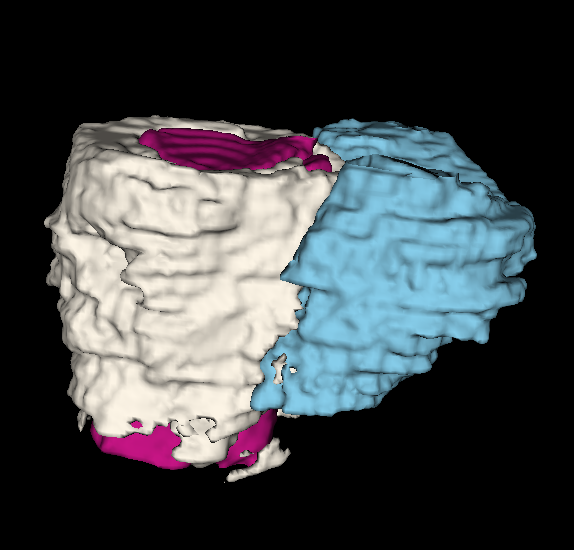}} &
		\hfill
		\subfloat[(d) pred.]{\adjincludegraphics[valign=c,height=1.1cm]{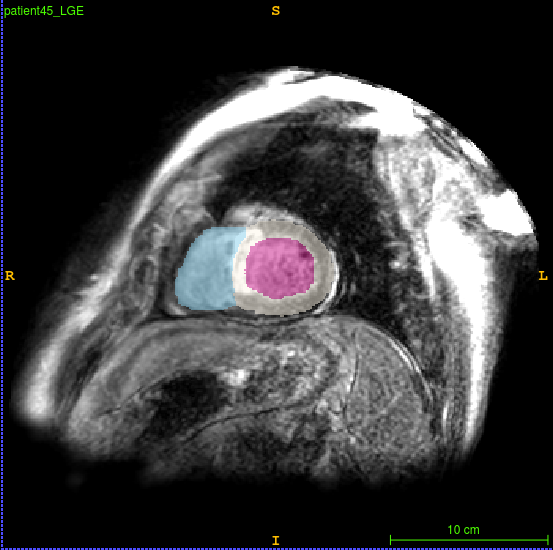}} &
		\hfill
		\subfloat[(e) pred. 3D]{\adjincludegraphics[valign=c,height=1.1cm]{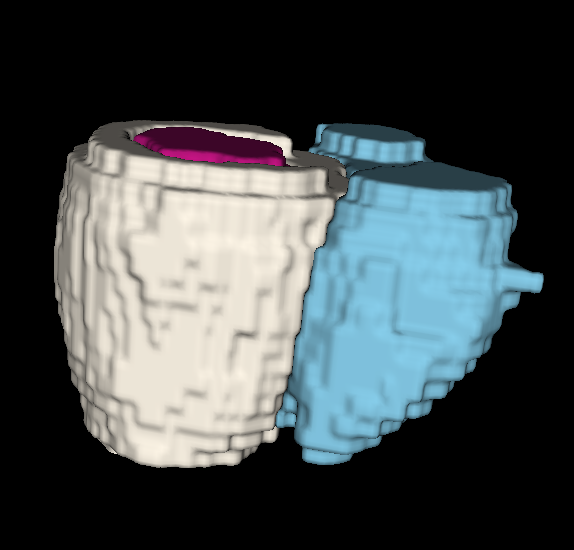}}
	\end{tabular}
	\caption{\footnotesize Comparison of the result after multi-atlas label fusion (b) and (c) in a testing LGE dataset (a) and our model's prediction (d) and (e) [patient 45 of the challenge dataset]. 
	\label{fig:pseudo_vs_model}}
\end{figure*}

In this work, we combined classical methods of multi-atlas label fusion with deep learning. We utilized the ability of multi-atlas label fusion to generate labels for new images using only a small set of labeled images of the targeted image modality as atlases, although resulting in less accurate (or ``noisy'') labels when compared to manual segmentation. Furthermore, we enhanced the noisy labels by merging more atlas-based label fusion results if annotations of the same patient's anatomy are available in different image modalities. Here, they came from different MRI sequences, but they could potentially stem from even more different modalities like CT, using multi-modality similarity measures to drive the registrations.
After training a round of deep convolutional neural networks on the ``noisy'' labels, we can see a clear visual improvement over multi-atlas label fusion result. This points to the fact that neural networks can still learn correlations of the data and the desired labels even when training labels are not as accurate as ground truth supervision labels \cite{heller2018imperfect}. The networks are able to compensate for some of the non-systematic errors in the ``noisy'' labels and hence improve the overall segmentation. We are blinded to the test set ground truth annotations and cannot quantify these improvements but visually, the improvements are noticeable as shown in Fig. \ref{fig:pseudo_vs_model}.
In conclusion, we achieved the automatic segmentation of cardiac structures in LGE magnetic resonance images by combing classical methods from multi-atlas label fusion and modern deep learning-based segmentation, resulting in visually compelling segmentation results.
%
%
\small
\bibliographystyle{splncs}

\begin{thebibliography}{10}

\bibitem{zhuang2018multivariate}
Zhuang, X.:
\newblock Multivariate mixture model for myocardial segmentation combining
  multi-source images.
\newblock IEEE transactions on pattern analysis and machine intelligence (2018)

\bibitem{zhuang2016multivariate}
Zhuang, X.:
\newblock Multivariate mixture model for cardiac segmentation from
  multi-sequence mri.
\newblock In: MICCAI, Springer (2016)  581--588

\bibitem{cciccek20163d}
{\c{C}}i{\c{c}}ek, {\"O}., Abdulkadir, A., Lienkamp, S.S., Brox, T.,
  Ronneberger, O.:
\newblock 3d u-net: learning dense volumetric segmentation from sparse
  annotation.
\newblock In: MICCAI, Springer (2016)  424--432

\bibitem{milletari2016v}
Milletari, F., Navab, N., Ahmadi, S.A.:
\newblock V-net: Fully convolutional neural networks for volumetric medical
  image segmentation.
\newblock In: 3D Vision (3DV), 2016 Fourth International Conference on, IEEE
  (2016)  565--571

\bibitem{myronenko20183d}
Myronenko, A.:
\newblock 3d mri brain tumor segmentation using autoencoder regularization.
\newblock In: International MICCAI Brainlesion Workshop, Springer (2018)
  311--320

\bibitem{zhu2019anatomynet}
Zhu, W., Huang, Y., Zeng, L., Chen, X., Liu, Y., Qian, Z., Du, N., Fan, W.,
  Xie, X.:
\newblock Anatomynet: Deep learning for fast and fully automated whole-volume
  segmentation of head and neck anatomy.
\newblock Medical physics \textbf{46}(2) (2019)  576--589

\bibitem{iglesias2015multi}
Iglesias, J.E., Sabuncu, M.R.:
\newblock Multi-atlas segmentation of biomedical images: a survey.
\newblock Medical image analysis \textbf{24}(1) (2015)  205--219

\bibitem{Rueckert1999nonrigid}
Rueckert, D., Sonoda, L., Hayes, C., Hill, D., Leach, M., Hawkes, D.:
\newblock Nonrigid registration using free-form deformations: Application to
  breast mr images.
\newblock IEEE Trans. Med. Imaging \textbf{18}(8) (1999)  712--721

\bibitem{modat2010fast}
Modat, M., Ridgway, G., Taylor, Z., Lehmann, M., Barnes, J., Hawkes, D., Fox,
  N., Ourselin, S.:
\newblock {Fast free-form deformation using graphics processing units}.
\newblock Comput. Meth. Prog. Bio. \textbf{98}(3) (2010)  278--284

\bibitem{studholme1999overlap}
Studholme, C., Hill, D.L., Hawkes, D.J.:
\newblock An overlap invariant entropy measure of 3d medical image alignment.
\newblock Pattern recognition \textbf{32}(1) (1999)  71--86

\bibitem{iglesias2015atlas}
Iglesias, J.E., Sabuncu, M.R.:
\newblock Multi-atlas segmentation of biomedical images: A survey.
\newblock Medical Image Analysis \textbf{24}(1) (2015)  205 -- 219

\bibitem{mattes2003pet}
Mattes, D., Haynor, D.R., Vesselle, H., Lewellen, T.K., Eubank, W.:
\newblock Pet-ct image registration in the chest using free-form deformations.
\newblock IEEE transactions on medical imaging \textbf{22}(1) (2003)  120--128

\bibitem{feng2009new}
Feng, W., Reeves, S., Denney, T., Lloyd, S., Dell'Italia, L., Gupta, H.:
\newblock A new consistent image registration formulation with a b-spline
  deformation model.
\newblock In: ISBI. (2009)  979--982

\bibitem{modat2012inverse}
Modat, M., Cardoso, M., Daga, P., Cash, D., Fox, N., Ourselin, S.:
\newblock Inverse-consistent symmetric free form deformation.
\newblock Biomedical Image Registration \textbf{7359} (2012)  79--88

\bibitem{liu20183d}
Liu, S., Xu, D., Zhou, S.K., Pauly, O., Grbic, S., Mertelmeier, T., Wicklein,
  J., Jerebko, A., Cai, W., Comaniciu, D.:
\newblock {3D} anisotropic hybrid network: Transferring convolutional features
  from {2D} images to {3D} anisotropic volumes.
\newblock In: MICCAI, Springer (2018)  851--858

\bibitem{he2016deep}
He, K., Zhang, X., Ren, S., Sun, J.:
\newblock Deep residual learning for image recognition.
\newblock In: CVPR. (2016)  770--778

\bibitem{heller2018imperfect}
Heller, N., Dean, J., Papanikolopoulos, N.:
\newblock Imperfect segmentation labels: How much do they matter?
\newblock In: Intravascular Imaging and Computer Assisted Stenting and
  Large-Scale Annotation of Biomedical Data and Expert Label Synthesis.
\newblock Springer (2018)  112--120

\end{thebibliography}

\end{document}